\documentstyle[preprint,aps]{revtex}

\begin{document}

\draft


\tighten

\title{Understanding Radiatively Induced Lorentz-CPT Violation 
  in Differential Regularization}

\author{ W.F. Chen\renewcommand{\thefootnote}{\dagger}\footnote{
E-mail: wchen@theory.uwinnipeg.ca}}
 
\address{ Department of Physics, University of Winnipeg, Winnipeg, 
Manitoba, R3B 2E9 Canada \\
   and\\
Winnipeg Institute for Theoretical Physics, Winnipeg, Manitoba}

\maketitle

\begin{abstract}
 We have investigated the perturbative ambiguity of the
radiatively induced Chern-Simons term in differential regularization.
The result obtained in this method contains 
all those obtained in other regularization schemes and the ambiguity is
explicitly characterized by an indefinite ratio of two renormalization
scales. It is argued that the ambiguity can only be eliminated
by either imposing a physical requirement or resorting to a 
more fundamental principle. Some calculation techniques in 
coordinate space are developed in the appendices. 

\vspace{3ex}
{\noindent PACS: 11.10.Kk; 11.15.-q}\\
Keywords: Differential regularization; Lorentz and CPT violation;
Regularization ambiguity; Convolution integral.
\end{abstract}

\vspace{3ex}

In a relativistic quantum field theory, Lorentz and CPT violating terms 
should be strictly prohibited. Otherwise, the status of
special relativity, as one of the cornerstone of modern physics, 
will be challenged. However,
recently some investigations have been carried out to consider this
possibility$\cite{ref1,ref2}$. 
Motivated by the proposal put forward about a decade ago
of introducing a Chern-Simons term$\cite{ref3}$, 
${\cal L}_k=1/2 k_{\mu}\epsilon^{\mu\nu\lambda\rho}F_{\nu\lambda}A_{\rho}$,
 to violate the Lorentz and CPT symmetry
of quantum electrodynamics, recently a Lorentz and CPT violating extension
of the standard model was constructed and some of its quantum aspects
were investigated$\cite{ref2}$. As pointed out by Jackiw$\cite{ref4}$, 
the availability of higher precision instruments
nowadays allows a more strict test on some of 
fundamental principles to be carried out. Such an 
investigation at least at theoretical level is not completely 
unreasonable. The question is how this Lorentz and CPT
violating  term can naturally arise rather than introducing it by hand.

Based on the experience in 2+1-dimensional QED$\cite{ref5}$, 
where a parity-odd Chern-Simons term
is induced from the fermionic determinant$\cite{ref6,ref7}$, 
one natural guess is that  
this Lorentz and CPT violation term can come from a
Lorentz and CPT violation term
$\bar{\psi}b\hspace{-2mm}/\gamma_5\psi$ in the fermionic sector. 
The explicit calculation carried out recently shows that 
this case can happen$\cite{ref8,ref9}$, there has induced a
Chern-Simons term with its coefficient $k_\mu$ proportional to 
$b_\mu$. However,  since the UV 
divergence usually emerges in a perturbative
 quantum correction, one must first choose a regularization
scheme to make the theory well defined. It was shown that the coefficient of 
this radiatively induced Chern-Simons term is regularization dependent
$\cite{ref2,ref8}$. In Pauli-Villars regularization, 
this coefficient is zero, while dimensional
regularization combined with the derivative expansion leads to a definite
non-zero value$\cite{ref8}$. It also seems to us that this induced term 
cannot be observed by calculating the fermionic determinant 
in Fock-Schwinger proper time method$\cite{ref10}$. 
In particular, based on the hypothesis that
the axial vector current $j_\mu^5=\bar{\psi}(x)\gamma_\mu\gamma_5\psi(x)$
should be gauge invariant at arbitrary four-momentum,
Coleman and Glashow claimed that the $k_{\mu}$ must be zero$\cite{ref11}$. 
Thus the existence of this radiatively induced Chern-Simons
term in perturbative theory is somehow ambiguous. More recently, 
a new calculation was performed by Jackiw 
and V.A. Kosteleck\'{y}$\cite{ref9}$, 
using the $b_\mu$-exact propagator instead of the free 
fermionic propagator, in which the Lorentz and CPT 
violating fermionic term  was treated non-perturbatively. 
It was shown that in this 
non-perturbative approach the Lorentz and CPT violating term 
is generated unambiguously at low-energy. 
Therefore, it is interesting to understand the 
discrepancies among these various results within the framework of 
perturbative theory.

 One possible way  is utilizing an improved regularization scheme.
It is known that a regularization is a 
temporary modification of the original theory. 
Different regularization 
schemes have actually provided different methods to 
calculate a quantum correction. 
Thus it is possible to incur a regularization dependent result.  To
avoid this occurrence, one should choose a regularization scheme
that modifies the original theory as little as possible
and preserves the features of the original theory such as symmetries
etc as much as possible. In view of this 
criteria, differential regularization 
seems to be the most appropriate candidate$\cite{ref12}$.
This regularization scheme is a relatively new calculation method and it
works for an Euclidean field theory in coordinate space. The invention
of this regularization is based on the observation that in coordinate
space the UV divergence manifest itself in the singularity preventing
the amplitude from having a Fourier transform into 
momentum space. So one can 
regulate the amplitude by writing its singular term
as the derivative of another less 
singular function, which has a well defined Fourier transform, then
performing  the Fourier transform and discarding the surface term.
In this way one can directly get a well defined amplitude.
Up to now this method and its modified version have been applied successfully
to almost every aspects of field theories, including chiral anomaly,
 low-dimensional and supersymmetric 
field theories$\cite{ref13,ref13a1,ref13a2,ref13a3,ref13a4}$.
One can easily see that this regularization method actually has never 
introduced an regulator to modify the Lagrangian of the original theory, 
 hence it does not pull the value of a primitively divergent Feynman 
diagram away from its singularity. In comparison with the
usual route of calculating a quantum correction, 
this method has actually skipped over the regularization  procedure 
and  straightforwardly yielded the renormalized 
result. Therefore, the quantum correction obtained in this regularization 
method should be more universal than any other 
regularization schemes and hence 
can provide a better understanding to above ambiguity. 

Not only these favourable features, differential regularization has 
another great advantage over other regularization schemes.  
When implementing a differential regularization on
a quantum amplitude, one can introduce a renormalization 
scale for each singular term. These individual renormalization scales
are not independent and the relations among them can be fixed by 
the symmetries of theory. In other words, the maintenance of 
the symmetries in the theory such as gauge symmetry etc can be 
achieved by choosing the indefinite
renormalization scales at the final stage of the calculation. 
As will be shown later, this special feature of differential
regularization is not only the reason why the regularization ambiguity
can be explicitly parameterized by the ratio of two renormalization scales,
but also provide a guide for us to search for a natural setting to 
eliminate this ambiguity.

In view of this, in this paper we shall investigate this radiatively induced
ambiguity in terms of differential regularization.
 The model we shall start from is quantum electrodynamics
with the inclusion of a Lorentz- and CPT- violating axial vector 
term in the fermionic sector$\cite{ref8,ref9,ref14}$,
\begin{eqnarray}
{\cal L}_{\rm fermion}=\bar{\psi}\left(\partial\hspace{-2.3mm}/
-A\hspace{-2.2mm}/-b\hspace{-1.9mm}/\gamma_5\right)\psi,
\end{eqnarray}
where $b_\mu$ is a constant four-vector with a fixed orientation in 
space-time. The term $\bar{\psi}b\hspace{-2mm}/\gamma_5$ is gauge
invariant, but it explicitly violates Lorentz- and CPT symmetries,
since $b_{\mu}$ picks up  a preferred direction in space-time.
We will see that this Lorentz- and CPT- violation in the fermionic
sector is the origin of the induced Chern-Simons term.

In Ref.$\cite{ref9}$, it was found that the radiatively
induced Chern-Simons term can arise in the low-energy, or
equivalently in the large fermionic mass limit. In principle, we can also 
utilize the $b_\mu$-exact propagator in coordinate space to calculate
the vacuum polarization tensor. However, the existence of the
$b$-term make it impossible to write out this $b_\mu$-exact propagator
in coordinate space, and hence one cannot proceed parallel to
 Ref.$\cite{ref5}$ in coordinate space. Thus we have to adopt 
a free fermionic propagator.
The Feynman diagram that will be calculated is 
the vacuum polarization tensor but with an
insertion of a zero-momentum composite operator 
$\int d^4z\bar{\psi}b\hspace{-2mm}/\gamma_5\psi$
in the internal fermionic line (Fig.1), since only this kind of 
diagram can give the lowest
order contribution in $b$ and hence possibly leads to the induced 
Chern-Simons term. Equivalently, this kind of Feynman diagram can also 
be thought as the triangle diagram composed of two vector currents
and one axial vector currents but with zero momentum transfer between
the vector currents.  
In fact, the explicit calculation in Ref.$\cite{ref9}$ 
is very similar to that for the chiral anomaly, 
only the zero-momentum transfer
between two vector gauge field vertices was achieved naturally
due to the utilization of the $b_\mu$-exact propagator. Here we can also get
a natural zero-momentum transfer by considering above Feynman diagrams. 

 We need the free fermionic propagator,
\begin{eqnarray}
S(x)=\frac{1}{4\pi^2}\partial\hspace{-2.3mm}/\frac{1}{x^2}
\end{eqnarray}
for the massless case, and 
 \begin{eqnarray}
S(x)=(\partial\hspace{-2.4mm}/-m)\Delta (x)=
\frac{m}{4\pi^2}(\partial\hspace{-2mm}/-m)\left[\frac{K_1(mx)}{x}\right]
\label{eq4}
\end{eqnarray}
for the massive case, here and later on we denote 
$x{\equiv}|x|$, $K_1(x)$ is 
the first-order modified Bessel function of the second kind. 
The short-distance expansion of the massive scalar 
propagator $\Delta (x)$ is 
\begin{eqnarray}
\Delta (x)&=&\frac{1}{4\pi^2}\frac{m}{x}
K_1(mx) \nonumber\\
&=&\frac{1}{4\pi^2}\left[\frac{1}{x^2}
+\frac{1}{2}m^2\ln (mx)+\frac{m^2}{4}\left(1-2\psi (2)\right)+
\mbox{regular terms}\right].
\label{eq4a}
\end{eqnarray}

We first have look at the massless case.   
The vacuum polarization with  an insertion of the zero-momentum composite 
operator $\int d^4z\bar{\psi}b\hspace{-2mm}/\gamma_5\psi$ on either of 
the fermionic lines is read  as
\begin{eqnarray}
\Pi_{\mu\nu}(x,y)&=&-b_{\lambda}\int d^4z\left\{\mbox{Tr}\left[\gamma_5\gamma_{\lambda}
S(z-x)\gamma_\mu S(x-y)\gamma_\nu S(y-z)\right]\right.\nonumber\\
&&\left.+\mbox{Tr}\left[\gamma_5\gamma_{\lambda}
S(z-y)\gamma_\nu S(y-x)\gamma_\mu S(x-z)\right]\right\}\nonumber\\
&=&-\frac{1}{(4\pi^2)^3}b_{\lambda}\left[\mbox{Tr}\left(\gamma_5\gamma_\lambda
\gamma_a\gamma_\mu\gamma_b\gamma_\nu\gamma_c\right)\int d^4z 
\frac{\partial}{\partial z_a}\frac{1}{(z-x)^2}
\frac{\partial}{\partial x_b}\frac{1}{(x-y)^2}
\frac{\partial}{\partial y_c}\frac{1}{(y-z)^2}\right.\nonumber\\
&&+\left.\mbox{Tr}\left(\gamma_5\gamma_\lambda
\gamma_a\gamma_\nu\gamma_b\gamma_\mu\gamma_c\right)\int d^4z 
\frac{\partial}{\partial z_a}\frac{1}{(z-y)^2}
\frac{\partial}{\partial y_b}\frac{1}{(y-x)^2}
\frac{\partial}{\partial x_c}\frac{1}{(x-z)^2}\right]\nonumber\\
&=&\frac{4}{(4\pi^2)^3}b_{\lambda}\left[\mbox{Tr}\left(\gamma_5\gamma_\lambda
\gamma_a\gamma_\mu\gamma_b\gamma_\nu\gamma_c\right)\int d^4z
\frac{(z-x)_a}{(z-x)^4}\frac{(z-y)_c}{(z-y)^4}
\frac{\partial}{\partial x_b}\frac{1}{(x-y)^2}\right.\nonumber\\
&&+\left.\mbox{Tr}\left(\gamma_5\gamma_\lambda
\gamma_a\gamma_\nu\gamma_b\gamma_\mu\gamma_c\right)\int d^4z
\frac{(z-y)_a}{(z-y)^4}\frac{(z-x)_c}{(z-x)^4}
\frac{\partial}{\partial y_b}\frac{1}{(x-y)^2}\right].
\end{eqnarray}
Using the convolution integral given by (\ref{eqa5}),
\begin{eqnarray}
\int d^4z\frac{(z-x)_{\mu}(z-y)_{\nu}}{(z-x)^4(z-y)^4}=
\frac{\pi^2}{2(x-y)^2}\left[\delta_{\mu\nu}-2\frac{(x-y)_{\mu}(x-y)_{\nu}}
{(x-y)^2}\right],
\end{eqnarray}
we obtain 
\begin{eqnarray}
\Pi_{\mu\nu}(x,y)&=&\Pi_{\mu\nu}(x-y)
=\frac{1}{32\pi^4}b_{\lambda}\mbox{Tr}\left[\gamma_5\gamma_\lambda
\left(\gamma_a\gamma_\mu\gamma_b\gamma_\nu\gamma_c-\gamma_a\gamma_\nu
\gamma_b\gamma_\mu\gamma_c\right)\right]\nonumber\\
&&{\times}\left[\delta_{ac}-
2\frac{(x-y)_{a}(x-y)_{c}}{(x-y)^2}\right]\frac{1}{(x-y)^2}
\frac{\partial}{\partial x_b}\frac{1}{(x-y)^2}.
\end{eqnarray}
For convenience, denoting $x-y$ as $x$ and employing the differential 
operation,
\begin{eqnarray}
\frac{x_a x_b x_c}{x^8}=
-\frac{1}{48}\frac{\partial^3}{\partial x_a \partial x_b \partial x_c}
\frac{1}{x^2}-\frac{1}{24}\left(\delta_{ab}\frac{\partial}{\partial x_c}
+\delta_{bc}\frac{\partial}{\partial x_a}
+\delta_{ca}\frac{\partial}{\partial x_b}\right)\frac{1}{x^4},
\end{eqnarray}
we can write the above vacuum polarization tensor as following form
\begin{eqnarray}
\Pi_{\mu\nu}(x)
&=&\frac{1}{64\pi^4}b_{\lambda}\mbox{Tr}\left[\gamma_5\gamma_\lambda
\left(\gamma_a\gamma_\mu\gamma_b\gamma_\nu\gamma_c-\gamma_a\gamma_\nu
\gamma_b\gamma_\mu\gamma_c\right)\right]\nonumber\\
&&{\times}\left[\frac{\partial}{\partial x_b}\frac{1}{x^4}\delta_{ac}
-\frac{1}{6}\frac{\partial^3}{\partial x_a \partial x_b \partial x_c}
\frac{1}{x^2}-\frac{1}{3}\left(\delta_{ab}\frac{\partial}{\partial x_c}
+\delta_{bc}\frac{\partial}{\partial x_a}
+\delta_{ca}\frac{\partial}{\partial x_b}\right)\frac{1}{x^4}\right].
\label{eq9}
\end{eqnarray}
Obviously, $1/x^4$ is too singular to have a Fourier transform into 
momentum space, so we must replace it by its differential regulated version
\begin{eqnarray}
\left(\frac{1}{x^4}\right)_R=-\frac{1}{4}\Box\frac{\ln (x^2M^2)}{x^2},
\label{eq10}
\end{eqnarray}
where $\Box{\equiv}\partial^2$ denotes the four-dimensional Laplacian operator.
Thus the differential regulated version of the vacuum polarization tensor 
with an insertion of the zero-momentum composite operator
 $\int d^4z\bar{\psi}b\hspace{-2mm}/\gamma_5\psi$ is
\begin{eqnarray}
\Pi_{\mu\nu}(x)
&=&\frac{1}{64\pi^4}b_{\lambda}\mbox{Tr}\left[\gamma_5\gamma_\lambda
\left(\gamma_a\gamma_\mu\gamma_b\gamma_\nu\gamma_c-\gamma_a\gamma_\nu
\gamma_b\gamma_\mu\gamma_c\right)\right]\nonumber\\
&&{\times}\left[-\frac{1}{4}\delta_{ca}\frac{\partial}{\partial x_b}
\Box\frac{\ln (x^2M_1^2)}{x^2}-\frac{1}{6}\frac{\partial^3}
{\partial x_a \partial x_b \partial x_c}\frac{1}{x^2}\right.\nonumber\\
&&\left.+\frac{1}{12}\left(\delta_{ab}\frac{\partial}{\partial x_c}
+\delta_{bc}\frac{\partial}{\partial x_a}
+\delta_{ca}\frac{\partial}{\partial x_b}\right)
\Box\frac{\ln (x^2M_2^2)}{x^2}\right].
\label{eq11}
\end{eqnarray}
Note that we chosen two different renormalization scales for $1/x^4$
in the first and the third term of Eq.(\ref{eq9}) since 
two singular terms can differ from a finite quantity. After contracting
with the external $\gamma$-matrix trace, we get 
\begin{eqnarray}
\Pi_{\mu\nu}(x)&=&\frac{1}{16\pi^4}b_{\lambda}\epsilon_{\lambda\mu\nu a}
\frac{\partial}{\partial x_a}\left(\frac{1}{3}+\ln\frac{M_1^2}{M_2^2}\right)\Box\frac{1}{x^2}
\nonumber\\
&=&-\frac{1}{4\pi^2}\left(\frac{1}{3}+2\ln\frac{M_1}{M_2}\right) b_{\lambda}\epsilon_{\lambda\mu\nu a}
\frac{\partial}{\partial x_a}\delta^{(4)}(x).
\label{eq12}
\end{eqnarray}
In above calculation, we have used 
\begin{eqnarray}
&&\mbox{Tr}\left[\gamma_5\gamma_\lambda
\left(\gamma_a\gamma_\mu\gamma_b\gamma_\nu\gamma_c-\gamma_a\gamma_\nu
\gamma_b\gamma_\mu\gamma_c\right)\right]\delta_{ca}
\frac{\partial}{\partial x_b}=-16
\epsilon_{\lambda\mu\nu b}\frac{\partial}{\partial x_b};\nonumber\\
&& \mbox{Tr}\left[\gamma_5\gamma_\lambda
\left(\gamma_a\gamma_\mu\gamma_b\gamma_\nu\gamma_c-\gamma_a\gamma_\nu
\gamma_b\gamma_\mu\gamma_c\right)\right]\delta_{bc}
\frac{\partial}{\partial x_a}=-16\
\epsilon_{\lambda\mu\nu a}\frac{\partial}{\partial x_a};\nonumber\\
&& \mbox{Tr}\left[\gamma_5\gamma_\lambda
\left(\gamma_a\gamma_\mu\gamma_b\gamma_\nu\gamma_c-\gamma_a\gamma_\nu
\gamma_b\gamma_\mu\gamma_c\right)\right]\delta_{ab}
\frac{\partial}{\partial x_c}=-16\
\epsilon_{\lambda\mu\nu c}\frac{\partial}{\partial x_c};\nonumber\\
&& \mbox{Tr}\left[\gamma_5\gamma_\lambda
\left(\gamma_a\gamma_\mu\gamma_b\gamma_\nu\gamma_c-\gamma_a\gamma_\nu
\gamma_b\gamma_\mu\gamma_c\right)\right]
\frac{\partial^3}{\partial x_a \partial x_b \partial x_c}
=-8\epsilon_{\lambda\mu\nu a}
\frac{\partial}{\partial x_a}\Box.
\end{eqnarray}

One may wonder why we only adopt two mass scales in  Eq.(\ref{eq11})
for those four short-distance singular terms of Eq.(\ref{eq9}). Of course,
we can introduce four distinct mass scales, then there appears
\begin{eqnarray}
\Pi_{\mu\nu}(x)
&=&\frac{1}{64\pi^4}b_{\lambda}\mbox{Tr}\left[\gamma_5\gamma_\lambda
\left(\gamma_a\gamma_\mu\gamma_b\gamma_\nu\gamma_c-\gamma_a\gamma_\nu
\gamma_b\gamma_\mu\gamma_c\right)\right]\nonumber\\
&&{\times}\left[-\frac{1}{4}\delta_{ac}\frac{\partial}{\partial x_b}
\Box\frac{\ln (x^2M_1^2)}{x^2}-\frac{1}{6}\frac{\partial^3}
{\partial x_a \partial x_b \partial x_c}\frac{1}{x^2}\right.\nonumber\\
&&\left.+\frac{1}{12}\left(\delta_{ab}\frac{\partial}{\partial x_c}
\Box\frac{\ln (x^2M_3^2)}{x^2}
+\delta_{bc}\frac{\partial}{\partial x_a}\Box\frac{\ln (x^2M_4^2)}{x^2}
+\delta_{ca}\frac{\partial}{\partial x_b}
\Box\frac{\ln (x^2M_5^2)}{x^2}\right)\right]\nonumber\\
&=&-\frac{1}{4\pi^2}\left[\frac{1}{3}+2\ln\frac{M_1}{(M_3M_4M_5)^{1/3}}\right]
b_{\lambda}\epsilon_{\lambda\mu\nu a}
\frac{\partial}{\partial x_a}\delta^{(4)}(x).
\label{eq11a}
\end{eqnarray}
With the definition $M_2{\equiv}(M_3M_4M_5)^{1/3}$, we still obtain the 
same result as Eq.(\ref{eq12}). It can be easily checked that
any other differential operations on $1/x^4$ will lead to the same 
conclusion: the ambiguity is only relevant to two independent
mass scales and uniquely parameterized by their ratio, $\ln M_1/M_2$.
The physical renormalization conditions or symmetries will fix
this ambiguity.  

The vacuum polarization tensor (\ref{eq12}) shows that the 
following Lorentz- and CPT violated action are indeed induced,
\begin{eqnarray}
S_{\rm ind}=\frac{1}{8\pi^2}\left(\frac{1}{3}+2\ln\frac{M_1}{M_2}\right)
 \int d^4x \epsilon_{\mu\nu\lambda\rho} b_{\mu}A_{\nu}F_{\lambda\rho}.
\end{eqnarray}
It is remarkable that this radiatively induced Lagrangian has 
an ambiguity parameterized 
by an indefinite coefficient $\ln M_1/M_2$.
It is just the case recently pointed out by Jackiw that the radiative
correction is finite but undetermined$\cite{ref14}$.

The more interesting case is when the fermion is massive, 
where Jackiw and Kosteleck\'{y}$\cite{ref9}$
successfully escaped from the ``no-go" theorem proposed by Coleman and 
Glashow $\cite{ref11}$ and found the existence of the radiatively 
induced Chern-Simons term in a non-perturbative way, 
so this case has a direct physical relevance.
The corresponding vacuum polarization tensor is
\begin{eqnarray}
\Pi_{\mu\nu}(x-y)&=&-\left(\frac{m}{4\pi^2}\right)^3
b_{\lambda}\int d^4z \left\{\mbox{Tr}
\left[\gamma_5\gamma_\lambda \left(\gamma_a
\frac{\partial}{\partial z_{a}}-m\right)
\frac{K_1[m(z-x)]}{z-x}\gamma_\mu 
\left(\gamma_b\frac{\partial}{\partial x_{b}}-m\right)\right.
\right.\nonumber\\
&&{\times}\left.\left.\frac{K_1[m(x-y)]}{x-y}\gamma_\nu 
\left(\gamma_c\frac{\partial}
{\partial y_{c}}-m\right)\frac{K_1[m(y-z)]}{y-z}\right]
+\mbox{Tr}\left(\mu{\leftrightarrow}\nu, x{\leftrightarrow}y\right)
\right\}\nonumber\\
&=&-\left(\frac{m}{4\pi^2}\right)^3
b_{\lambda}\int d^4z \left\{\mbox{Tr}\left(\gamma_5\gamma_\lambda\gamma_a
\gamma_\mu\gamma_b\gamma_\nu\gamma_c\right)\frac{\partial}{\partial z_{a}}
\frac{K_1[m(z-x)]}{z-x}
\right.\nonumber\\
&&{\times}\frac{\partial}{\partial x_{b}}\frac{K_1[m(x-y)]}{x-y}
\frac{\partial}{\partial y_{c}}\frac{K_1[m(y-z)]}{y-z}
\nonumber\\
&+&\mbox{Tr}\left(\gamma_5\gamma_\lambda\gamma_a
\gamma_\nu\gamma_b\gamma_\mu\gamma_c\right)\frac{\partial}{\partial z_{a}}
\frac{K_1[m(z-y)]}{z-y}\frac{\partial}{\partial y_{b}}
\frac{K_1[m(y-x)]}{y-x}
\frac{\partial}{\partial x_{c}}\frac{K_1[m(x-z)]}{x-z}\nonumber\\
&+&8 m^2\epsilon_{\lambda\mu\nu a}\left[\left(\frac{\partial}{\partial z_{a}}
\frac{K_1[m(z-x)]}{z-x}\right) \frac{K_1[m(x-y)]}{x-y}
\frac{K_1[m(y-z)]}{y-z}\right.\nonumber\\
&&+\frac{K_1[m(z-y)]}{z-y}
\left(\frac{\partial}{\partial y_{a}}\frac{K_1[m(y-x)]}{y-x}\right)
\frac{K_1[m(x-z)]}{x-z}\nonumber\\
&&\left.\left.+\frac{K_1[m(z-x)]}{z-x}\frac{K_1[m(x-y)]}{x-y}
\left(\frac{\partial}{\partial y_{a}}\frac{K_1[m(y-z)]}{y-z}\right)\right]
\right\}.
\label{eq15}
\end{eqnarray}
Using the convolution integral (\ref{eqa13}) for the massive case 
and denoting $x-y$ as $x$, we can write the vacuum polarization 
tensor (\ref{eq15}) as follows
\begin{eqnarray}
\Pi_{\mu\nu}(x)&=&-\left(\frac{m}{4\pi^2}\right)^3
b_{\lambda}\left\{\frac{2\pi^2}{m^2}
\left[\mbox{Tr}\left(\gamma_5\gamma_\lambda\gamma_a
\gamma_\mu\gamma_b\gamma_\nu\gamma_c\right)
-\mbox{Tr}\left(\gamma_5\gamma_\lambda\gamma_a
\gamma_\nu\gamma_b\gamma_\mu\gamma_c\right)\right]\right.\nonumber\\
&&\times \frac{\partial}{\partial x_b}\frac{K_1(mx)}{x}
\frac{\partial^2}{\partial x_a \partial x_c}K_0(mx)\nonumber\\
&-&\left.\frac{2\pi^2}{m^2} 
\epsilon_{\lambda\mu\nu a}\left[16m^2 \frac{K_1(mx)}{x}
\frac{\partial}{\partial x_a}K_0(m x)
+8 m^2 K_0(mx)\frac{\partial}{\partial x_a} \frac{K_1(mx)}{x}\right]\right\}.
\label{eq16}
\end{eqnarray}
One natural way to perform the operation on (\ref{eq16}) is 
to expand the term $\partial/\partial x_b$$(K_1(mx)/x)$
$\partial^2/(\partial x_a\partial x_c)$$
K_0(mx)$, write its singular terms in a derivative form and then
contract it with $\gamma$-matrix trace. However,
we have no way to realize this due to the difficulty in solving
a differential equation with the modified Bessel function. 
Neither can  we do it for the terms
$K_1(mx)/x \partial/\partial x_a K_0(mx)$ and 
$K_0(mx)\partial/\partial x_a K_0(mx)$. Therefore, in contrast to
the massless case, we shall first carry out the trace calculation.
Making use of the techniques collected in (\ref{eqb1})-(\ref{eqb4}), 
we can work out above vacuum polarization tensor as follows,
\begin{eqnarray}
\Pi_{\mu\nu}(x)&=&-\frac{m}{32\pi^4}b_{\lambda}
\left\{16\epsilon_{\lambda a b\nu} \frac{\partial}{\partial x_a}
\left[\frac{\partial}{\partial x_b}\frac{K_1(mx)}{x}
\frac{\partial}{\partial x_\mu}\frac{K_0(mx)}{x}\right]\right.\nonumber\\
&-&16\epsilon_{\lambda a b\mu } \frac{\partial}{\partial x_a}
\left[\frac{\partial}{\partial x_b}\frac{K_1(mx)}{x}
\frac{\partial}{\partial x_\nu}\frac{K_0(mx)}{x}\right]\nonumber\\
&-&16\epsilon_{\lambda\mu\nu a} \frac{\partial}{\partial x_b} 
\frac{K_1(mx)}{x} 
\frac{\partial^2}{\partial x_a \partial x_b}K_0(mx)
+8\epsilon_{\lambda\mu\nu a}\frac{\partial}{\partial x_a} \frac{K_1(mx)}{x}
\Box K_0(mx)\nonumber\\
&-& \left. 8 m^2 \epsilon_{\lambda\mu\nu a}\left[ 2 \frac{K_1(mx)}{x}
\frac{\partial}{\partial x_a}K_0(m x)
+ K_0(mx)\frac{\partial}{\partial x_a} \frac{K_1(mx)}{x}\right]\right\}\nonumber\\
&=&\frac{m^2}{2\pi^4}b_{\lambda}\epsilon_{\lambda\mu\nu a}\left\{
-2\left[\frac{\partial}{\partial x_a}\left(\frac{K_1(mx)}{x}\right)^2
- \frac{\partial}{\partial x_a}\left(\frac{K_1(mx)}{x}\right)^2\right]
\right.\nonumber\\
&&\left.+m\frac{\partial}{\partial x_a}\left[\frac{K_1(mx)K_0(mx)}{x}\right]\right\}.
\label{eq21}
\end{eqnarray} 
Obviously, due to the asymptotic expansion (\ref{eq4a})
the function $[K_1(mx)/x]^2$ is singular as $x{\sim}0$ and 
has no Fourier transform into the momentum space.
It should be emphasized that in deriving 
Eqs.(\ref{eqb2})--(\ref{eqb4}) and (\ref{eq21})
the substraction operation among the singular terms like $[K_1(mx)/x]^2$
should not be naively carried out. It is analogous to the fact
that in momentum space two divergent terms
with the same form but opposite sign cannot be canceled,
until after a regularization scheme is implemented so that they become
well defined and the substraction operation can work safely. Otherwise,
a finite term will probably be lost since in general the difference
of two infinite quantities is not zero. 
In fact, the operation keeping the singular terms untouched before
performing the regularization is a crucial point in the differential 
regularization method.  
 
Unfortunately, as above, due to the difficulty in solving 
a differential equation with the modified Bessel function, 
we still cannot write
the singular function $[K_1(mx)/x]^2$ as the derivative of another
less singular function. However, we can consider the asymptotic expansion 
(\ref{eq4a}). One can easily see that in Eq.(\ref{eq21}) the singularity 
at short-distance is only carried by the leading term $1/x^4$, 
the other terms are finite and hence they are exactly canceled. 
Therefore, making use of Eq.(\ref{eq10}) again,
we obtained the regulated form for the vacuum polarization tensor
in the massive case,
\begin{eqnarray}
\Pi_{\mu\nu}(x)&=&\frac{m^2}{2\pi^4}b_{\lambda}\epsilon_{\lambda\mu\nu a}
\left\{-\frac{1}{2m^2}\frac{\partial}{\partial x_a}\left[
\Box\frac{\ln x^2M^2_1}{x^2}-
\Box\frac{\ln x^2M^2_2}{x^2}\right]+m\frac{\partial}{\partial x_a}
\left[\frac{K_1(mx)K_0(mx)}{x}\right]\right\}\nonumber\\
&=&\frac{1}{2\pi^4}b_{\lambda}\epsilon_{\lambda\mu\nu a}
\left\{-\ln\frac{M_1}{M_2}\frac{\partial}{\partial x_a}
\Box\frac{1}{x^2}+m^3\frac{\partial}{\partial x_a}
\left[\frac{K_1(mx)K_0(mx)}{x}\right]\right\}
\nonumber\\
&=&\frac{1}{2\pi^4}b_{\lambda}\epsilon_{\lambda\mu\nu a}
\left\{4\pi^2 \ln\frac{M_1}{M_2}\frac{\partial}{\partial x_a}\delta^{(4)}(x)
+ m^3\frac{\partial}{\partial x_a}
\left[\frac{K_1(mx)K_0(mx)}{x}\right]\right\}.
\label{eq23}
\end{eqnarray} 
The above vacuum polarization tensor can be expressed in momentum space
by performing its Fourier transform. According to the standard differential
regularization procedure$\cite{ref12}$, we have
\begin{eqnarray}
\Pi_{\mu\nu}(p)&=&\int d^4xe^{-ip{\cdot}x}\Pi_{\mu\nu}(x)\nonumber\\
&=&\frac{2}{\pi^2}b_{\lambda}\epsilon_{\lambda\mu\nu a}
ip_a\left[ \ln\frac{M_1}{M_2}+\frac{m^3}{4\pi^2}
\int d^4x e^{-ip{\cdot}x}
\frac{K_1(mx)K_0(mx)}{x}\right]\nonumber\\
&=&\frac{2}{\pi^2}b_{\lambda}\epsilon_{\lambda\mu\nu a}
ip_a\left[\ln\frac{M_1}{M_2}
+\frac{m}{2p}\frac{\arcsin\hspace{-0.8mm}\mbox{h}[p/(2m)]}
{\sqrt{1+p^2/(4m^2)}}\right]\nonumber\\
&=&\frac{2}{\pi^2}b_{\lambda}\epsilon_{\lambda\mu\nu a}ip_a
\left[\ln\frac{M_1}{M_2}+\frac{m}{4p\sqrt{1+p^2/(4m^2)}}
\ln\frac{\sqrt{1+p^2/(4m^2)}+p/(2m)}{\sqrt{1+p^2/(4m^2)}-p/(2m)}\right].
\label{eq24}
\end{eqnarray}
As Ref.$\cite{ref5}$, the radiatively induced Chern-Simons term
can be defined  at low-energy $p^2=0$ 
(or equivalently at large-$m$ limit)
\begin{eqnarray}
\Pi_{\lambda\mu\nu}(p)|_{p^2=0}=\frac{2}{\pi^2}
\epsilon_{\lambda\mu\nu a}ip_a
\left(\ln\frac{M_1}{M_2}+\frac{1}{4}\right).
\label{eq25}
\end{eqnarray}

Eq.(\ref{eq25}) shows that the coefficient of the 
induced Chern-Simons term has a finite ambiguity, which was explicitly
parameterized by the ratio of two renormalization scales, $M_1/M_2$.
Especially, Eq.(\ref{eq25}) has contained all the results obtained in other
regularization schemes. For $M_1=e^{-1/4}M_2$, we get the conclusion
in Pauli-Villars regularization,
\begin{eqnarray}
\Pi_{\lambda\mu\nu}(p)|_{p^2=0}=0;
\label{eq26}
\end{eqnarray}
While if we choose  $M_1=e^{-1/16}M_2$, then the result  in 
dimensional regularization$\cite{ref8}$ and the non-perturbative approach
$\cite{ref9}$ is reproduced,
\begin{eqnarray}
\Pi_{\lambda\mu\nu}(p)|_{p^2=0}
=\frac{3}{8\pi^2}\epsilon_{\lambda\mu\nu a}ip_a.
\label{eq27}
\end{eqnarray}
It is remarkable that a natural choice $M_1=M_2$ does not 
correspond to a subtraction in the dispersive representation
given in Ref.$\cite{ref9}$.

The above conclusion is not strange to us and the profound reason lies in 
the excellent features possessed by differential regularization. As it 
is shown above, the basic operation in differential regularization
is replacing a singular term by the derivative of another less singular 
function. This operation has provided a possibility to add
arbitrary local terms to the higher order amplitude since we have 
to solve a differential equation for non-coincident points$\cite{ref15}$. 
When performing such a operation, we are introducing a new 
arbitrary local term into the quantum effective action. 
According to renormalization theory, the introduction of 
an arbitrary local term into the amplitude of a Green function
is equivalent to the addition of a finite counterterm to the Lagrangian. 
Therefore, from this viewpoint, differential regularization can lead to
a more general quantum effective action than any other regularization 
schemes. In particular, differential regularization keeps all the
ambiguities to the final stage of the calculation, and these ambiguities
can only be fixed by imposing some additional physical requirements or 
resorting to some more fundamental principle. This special
feature presented by differential regularization has formed  a sharp
contrast to other regularization schemes such as dimensional,
Pauli-Villars and cut-off regularization etc. These regularization
methods, together with the given renormalization prescription, can fix the those
arbitrary terms automatically at the beginning. In different regularization
schemes, these local terms are different. This is the reason why 
different regularization schemes can induce different Chern-Simons terms.
It should be emphasized that no regularization can claim 
that it gives the right value for this induced term. 
In differential regularization, this ambiguity is explicitly 
parameterized by the ratio of two indefinite renormalization 
scales and the results obtained in other regularization schemes can 
be reproduced by an appropriate choices on this arbitrary ratio. 
Therefore, one can say that differential regularization
has yielded a more universal result than any other regularization method,
since it does not impose any preferred choice on the Green  function at 
the beginning.

  In summary, we have investigated the
Lorentz- and CPT violating Chern-Simons term induced by radiative
corrections in differential 
regularization. The ambiguous results obtained in other regularization 
schemes are universally obtained 
and especially, the ambiguity is quantitively parameterized by the
ratio of two renormalization scales. 
This ambiguity should be fixed by renormalization conditions or
certain fundamental physical symmetries rather than an arbitrary
choice on the mass scales. For example, if one requires the 
Lagrangian density rather than the action to be gauge invariant, one
must choose $M_1=e^{-1/4}M_2$, and hence the generated Chern-Simons term
vanishes. Another choice is, to require the action 
and $\int d^4x j_{\mu}^5$ (i.e. the axial vector current $j_{\mu}^5$$=\bar{\psi}\gamma_\mu\gamma_5\psi$ at zero momentum) 
to be gauge invariant, as done by Jackiw and Kosteleck\'{y}$\cite{ref9}$. 
In this case, one must choose $M_1=e^{-1/16}M_2$ and 
consequently, the Chern-Simons term is generated unambiguously.
It should be emphasized that the natural prescription on the renormalization
scales, $M_1=M_2$, can be taken only when it corresponds to certain
physical renormalization condition.

\acknowledgments

This work is supported by the Natural Sciences and Engineering 
Research Council of Canada. I am very grateful to Prof. R. Jackiw 
for his suggesting me this problem, the enlightening discussions 
and comments on the manuscript. I am greatly indebted to Prof. 
G. Kunstatter for his continuous discussions and improvement 
on this manuscript. I would like to thank Dr. M. Carrington 
and Prof. R. Kobes for their encouragements and help. I am also 
obliged to Dr. M. Perez-Victoria for his useful discussions
on differential regularization.

\appendix 

\setcounter{equation}{0}

\section{ derivation of convolution integral}

One important technique in our calculation is the application
of the convolution integrals. Here we give a detail derivation.

In the massless case, there exists that
\begin{eqnarray}
\int d^4z\frac{(z-x)_{\mu}(z-y)_{\nu}}{(z-x)^4(z-y)^4}
=\frac{1}{4}\frac{\partial}{\partial x_{\mu}}
\frac{\partial}{\partial y_{\nu}}\int d^4z \frac{1}{(z-x)^2(z-y)^2}.
\end{eqnarray}
We need to write the integration $\int 1/[(z-x)^2 (z-y)^2]$ as
an explicit function of $x-y$.  So we first assume 
\begin{eqnarray}
f(x-y)=\int d^4z \frac{1}{(z-x)^2(z-y)^2}.
\label{eqa2}
\end{eqnarray}
Acting the four-dimensional Laplacian operator $\Box_x$ on the both sides
of Eq.(\ref{eqa2}) and using the formula,
\begin{eqnarray}
\Box_x\frac{1}{x^2}=-4\pi^2 \delta^{(4)}(x),
\end{eqnarray}
we obtain
\begin{eqnarray}
\Box_x f(x-y)=-4\pi^2\frac{1}{(x-y)^2}.
\label{eqa4}
\end{eqnarray}
With aid of the (four-dimensional) spherical symmetric form  of the 
Laplacian operator $\Box_x=4/x^2 d/d x^2[(x^2)^2d/d x^2]$, 
the solution to the differential equation (\ref{eqa4}) can be easily found,
\begin{eqnarray}
f(x-y)=-\pi^2\ln[\Lambda^2 (x-y)^2],
\end{eqnarray}
$\Lambda$ being the cut-off.
Thus we obtain the convolution integral formula
\begin{eqnarray}
\int d^4z\frac{(z-x)_{\mu}(z-y)_{\nu}}{(z-x)^4(z-y)^4}
=\frac{\pi^2}{2}\left[\delta_{\mu\nu}
-2\frac{(x-y)_\mu (x-y)_\nu}{(x-y)^4}\right].
\label{eqa5}
\end{eqnarray}

Now we turn to the massive case, where the situation is
quite complicated. From Eq.(\ref{eq15}), 
what we need to determine is \begin{eqnarray}
g(x-y)=\int d^4z\frac{K_1[m(z-x)]}{z-x}\,\frac{K_1[m(z-y)]}{z-y}.
\label{eqa7}
\end{eqnarray}
According to the property
\begin{eqnarray}
(\Box-m^2)\Delta (x)=(\Box-m^2)\left[\frac{1}{4\pi^2}
\frac{m}{x}K_1(mx)\right]=\delta^{(4)}(x),
\end{eqnarray}
we act the operator $(\Box_x-m^2)$ on both sides of Eq.(\ref{eqa7})
and obtain
\begin{eqnarray}
(\Box_x-m^2) g(x-y)=4\pi^2 \frac{K_1[m(x-y)]}{m(x-y)}.
\label{eqa9}
\end{eqnarray}
Repeating the above operation on Eq.(\ref{eqa9}), we get
\begin{eqnarray}
(\Box_x-m^2)^2 g(x)=\frac{16\pi^4}{m^2}\delta^{(4)}(x).
\end{eqnarray}
Upon considering the Fourier transform of $g(x)$,
\begin{eqnarray}
g(x)=\int \frac{d^4p}{(2\pi)^4}\,g(p)e^{ip{\cdot}x},
\label{eqa10}
\end{eqnarray}
Eq.(\ref{eqa10}) directly yields
\begin{eqnarray}
g(p)=\frac{16\pi^4}{m^2} \,\frac{1}{(p^2+m^2)^2}.
\label{eqa11}
\end{eqnarray}
$g(x)$ can be obtained by performing the Fourier transform,
\begin{eqnarray}
 g(x)&=& \frac{16\pi^4}{m^2}\int \frac{d^4p}{(2\pi)^4}\, 
\frac{1}{(p^2+m^2)^2} e^{ip{\cdot}x}\nonumber\\
&=&\frac{1}{m^2}\int_0^\infty dp\frac{p^3}{(p^2+m^2)^2}
\int_0^{\pi}d\theta \sin^2\theta e^{ipx \cos\theta}\int_0^\pi \sin\varphi
d\varphi \int_0^{2\pi}d\phi\nonumber\\
&=&\frac{4\pi^2}{m^2 x}\int_0^\infty dp\frac{p^2}{(p^2+m^2)^2}J_1(px)
=\frac{2\pi^2}{m^2}K_0(m x).
\label{eqa12}
\end{eqnarray}  
Thus we have finally worked out the important convolution
integral formula,
\begin{eqnarray}
\int d^4z\frac{K_1[m(z-x)]}{z-x}\,\frac{K_1[m(z-y)]}{z-y}=
\frac{2\pi^2}{m^2}K_0[m(x-y)],
\label{eqa13}
\end{eqnarray} 
$K_0(x)$ being the zeroth-order modified Bessel function of the second kind.

\setcounter{equation}{0}

\section{Differential Operations}

Some differential calculation techniques used in deriving the massive 
vacuum polarization tensor (\ref{eq23}) is collected in this appendix:

\begin{eqnarray}
&&\mbox{Tr}\left(\gamma_5\gamma_\lambda\gamma_a
\gamma_\mu\gamma_b\gamma_\nu\gamma_c\right)
\frac{\partial}{\partial x_b}\frac{K_1(mx)}{x}
\frac{\partial^2}{\partial x_a \partial x_c}K_0(mx)\nonumber\\
&=&8\epsilon_{\lambda a\mu b} \frac{\partial}{\partial x_a}
\left[\frac{\partial}{\partial x_b}\frac{K_1(mx)}{x}
\frac{\partial}{\partial x_\nu}\frac{K_0(mx)}{x}\right]\nonumber\\
&+&8\epsilon_{\lambda a b\nu } \frac{\partial}{\partial x_a}
\left[\frac{\partial}{\partial x_b}\frac{K_1(mx)}{x}
\frac{\partial}{\partial x_\mu}\frac{K_0(mx)}{x}\right]\nonumber\\
&+&8\epsilon_{\lambda\mu\nu a} \frac{\partial}{\partial x_b} 
\frac{K_1(mx)}{x} \frac{\partial^2}{\partial x_a \partial x_b}K_0(mx)\nonumber\\
&+&4\epsilon_{\lambda\mu\nu a}\frac{\partial}{\partial x_a} \frac{K_1(mx)}{x}
\Box K_0(mx);
\label{eqb1}
\end{eqnarray}
\begin{eqnarray}
\epsilon_{\lambda a b\nu}\frac{\partial}{\partial x_a}\left[
\frac{\partial}{\partial x_b}\frac{K_1(mx)}{x}\frac{\partial}{\partial x_\mu}
K_0(mx)\right]
&=&\epsilon_{\lambda ab \nu}\frac{\partial}{\partial x_a}\left[-m x_{\mu}
\frac{K_1(mx)}{x}\frac{\partial}{\partial x_b}\frac{K_1(mx)}{x}\right] 
\nonumber\\
&=&-\frac{1}{2}m\epsilon_{\lambda a b\nu}\frac{\partial}{\partial x_a}
\left[x_{\mu}\frac{\partial}{\partial x_b}\left(\frac{K_1(mx)}{x}\right)^2
\right]\nonumber\\
&=&\frac{1}{2}m\epsilon_{\lambda\mu\nu b}\frac{\partial}{\partial x_b}\left(\frac{K_1(mx)}{x}\right)^2;
\label{eqb2}
\end{eqnarray}
\begin{eqnarray}
&&\epsilon_{\lambda\mu\nu a}\frac{\partial}{\partial x_b}\frac{K_1(mx)}{x}
\frac{\partial^2}{\partial x_a \partial x_b}K_0(mx)\nonumber\\
&=&-m\epsilon_{\lambda\mu\nu a}\frac{\partial}{\partial x_b}\frac{K_1(mx)}{x}
\left\{\delta_{ab}\frac{K_1(mx)}{x}
+x_ax_b\frac{1}{x}\frac{d}{dx}\left[\frac{K_1(mx)}{x}\right]\right\}\nonumber\\
&=&-m\epsilon_{\lambda\mu\nu a}\left[\frac{1}{2}
\frac{\partial}{\partial x_a}\left(\frac{K_1(mx)}{x}\right)^2
+\left(\frac{\partial}{\partial x_a}\frac{K_1(mx)}{x}\right)
x\frac{d}{dx} \frac{K_1(mx)}{x}\right]\nonumber\\
&=&-m\epsilon_{\lambda\mu\nu a}\left\{\frac{1}{2}
\frac{\partial}{\partial x_a}\left(\frac{K_1(mx)}{x}\right)^2
+\frac{\partial}{\partial x_a}\frac{K_1(mx)}{x}
\left[-\frac{K_1(mx)}{x}+\frac{d}{dx}K_1(mx)\right]
\right\}\nonumber\\
&=&-m\epsilon_{\lambda\mu\nu a}\left[
\frac{1}{2}
\frac{\partial}{\partial x_a}\left(\frac{K_1(mx)}{x}\right)^2
-\frac{1}{2}
\frac{\partial}{\partial x_a}\left(\frac{K_1(mx)}{x}\right)^2+
\frac{\partial}{\partial x_a}
\frac{K_1(mx)}{x}\frac{d}{dx}K_1(mx)\right]\nonumber\\
&=&-m\epsilon_{\lambda\mu\nu a}\left\{\frac{1}{2}
\frac{\partial}{\partial x_a}\left(\frac{K_1(mx)}{x}\right)^2
-\frac{1}{2}\frac{\partial}{\partial x_a}\left(\frac{K_1(mx)}{x}\right)^2
\right.\nonumber\\
&&\left.-\frac{\partial}{\partial x_a}
\frac{K_1(mx)}{x}\left[m K_0(mx)+\frac{K_1(mx)}{x}\right]\right\}
\nonumber\\
&=&\epsilon_{\lambda\mu\nu a}\left[-\frac{1}{2}m 
\frac{\partial}{\partial x_a}\left(\frac{K_1(mx)}{x}\right)^2
+m\frac{\partial}{\partial x_a}\left(\frac{K_1(mx)}{x}\right)^2+
m^2K_0(mx)\frac{\partial}{\partial x_a}
\frac{K_1(mx)}{x}\right];
\label{eqb3}
\end{eqnarray}
\begin{eqnarray}
&&\epsilon_{\lambda\mu\nu a}\frac{\partial}{\partial x_a}\frac{K_1(mx)}{x}
\Box K_0(mx)\nonumber\\
&=&-m\epsilon_{\lambda\mu\nu a}
\frac{\partial}{\partial x_a}\frac{K_1(mx)}{x}\left[
4\frac{K_1(mx)}{x}+x \frac{d}{dx}\frac{K_1(mx)}{x}\right]\nonumber\\
&=&\epsilon_{\lambda\mu\nu a}\left[-2m\frac{\partial}{\partial x_a}
\left(\frac{K_1(mx)}{x}\right)^2+m\frac{\partial}{\partial x_a}
\left(\frac{K_1(mx)}{x}\right)^2\right.\nonumber\\
&&+\left.m^2 K_0(mx)\frac{\partial}{\partial x_a}
\frac{K_1(mx)}{x}\right].
\label{eqb4}
\end{eqnarray}

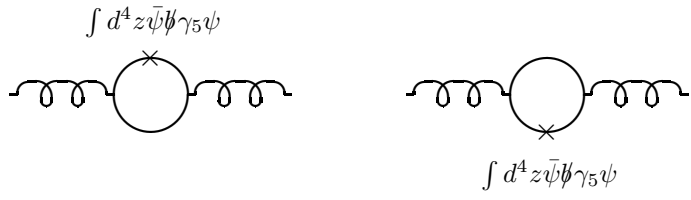
\begin{figure}
\centering
\input FEYNMAN
\begin{picture}(50000,10000)
\THICKLINES
\put(11500,7500){\footnotesize$\int d^4z\bar{\psi}b\hspace{-1.5mm}/\gamma_5\psi$}
\put(13500,6080){$\times$}
\put(14000,5000){\circle{3000}}
\startphantom
\drawline\gluon[\E\REG](0,0)[1]\gluoncap
\stopphantom
\pbackx=12600 \pbacky=5000
\global\multiply\plengthx by -1
\global\multiply\plengthy by -1
\global\advance\pbackx by \plengthx
\global\advance\pbacky by \plengthy
\drawline\gluon[\E\REG](\pbackx,\pbacky)[1]\gluoncap
\drawline\gluon[\W\FLIPPED](\gluonfrontx,\gluonfronty)[1]\gluoncap
\gluonbackx=15400 \gluonbacky=5000
\negate\gluonlengthx
\negate\gluonlengthy
\global\advance\gluonbackx by \gluonlengthx
\global\advance\gluonbacky by \gluonlengthy
\drawline\gluon[\W\FLIPPED](\gluonbackx,\gluonbacky)[1]\gluoncap
\drawline\gluon[\E\REG](\gluonfrontx,\gluonfronty)[1]\gluoncap

\put(26500,1700){\footnotesize$\int d^4z\bar{\psi}b\hspace{-1.5mm}/\gamma_5\psi$}
\put(28500,3280){$\times$}
\put(29000,5000){\circle{3000}}
\startphantom
\drawline\gluon[\E\REG](0,0)[1]\gluoncap
\stopphantom
\pbackx=27600 \pbacky=5000
\global\multiply\plengthx by -1
\global\multiply\plengthy by -1
\global\advance\pbackx by \plengthx
\global\advance\pbacky by \plengthy
\drawline\gluon[\E\REG](\pbackx,\pbacky)[1]\gluoncap
\drawline\gluon[\W\FLIPPED](\gluonfrontx,\gluonfronty)[1]\gluoncap
\gluonbackx=30400 \gluonbacky=5000
\negate\gluonlengthx
\negate\gluonlengthy
\global\advance\gluonbackx by \gluonlengthx
\global\advance\gluonbacky by \gluonlengthy
\drawline\gluon[\W\FLIPPED](\gluonbackx,\gluonbacky)[1]\gluoncap
\drawline\gluon[\E\REG](\gluonfrontx,\gluonfronty)[1]\gluoncap
\end{picture}
\caption{\protect\small Vacuum polarization contributed by
fermionic loop with an insertion of zero-momentum composite operator
$\int d^4z \bar{\psi}b\hspace{-1.6mm}/\gamma_5\psi$ in either of the internal
fermionic lines, $\times$ denoting the zero-momentum composite operator 
$\int d^4z \bar{\psi}b\hspace{-1.8mm}/\gamma_5\psi$.}

\end{figure}
\end{document}